\documentclass[a4paper,aps,prd,10pt,preprintnumbers,showpacs,twocolumn,superscriptaddress,nofootinbib,amsmath,amssymb]{revtex4-1}
\usepackage{graphicx}
\usepackage[utf8]{inputenc}
\usepackage[T1]{fontenc}
\usepackage{cmap}
\usepackage{ulem}

 \def\a{\alpha}\def\b{\beta}\def\g{\gamma}
  
  \def\om{\omega}

 \def\P{\Psi} 

\def\imo{i}

\def\be{\begin{equation}}
\def\ee{\end{equation}}
\def\bea{\begin{eqnarray}}
\def\eea{\end{eqnarray}}

\begin{document}

\title{Overtones of black holes via time-domain integration}
\author{Alexey Dubinsky}
\email{dubinsky@ukr.net}
\affiliation{University of Seville, 41009 Sevilla, Spain}

\begin{abstract}
We show that the first several overtones could be calculated by the time-domain integration method for asymptotically de Sitter black holes already at the lowest multipole numbers of gravitational and electromagnetic perturbations. This is not possible for asymptotically flat black holes, for which extraction of frequencies with the Prony method is usually possible with reasonable accuracy only for the fundamental mode. The reason for much better efficiency in the de Sitter case is the absence of power-law tails: the quasinormal modes dominate the signal not only at the intermediate stage, but also at asymptotically late times.
\end{abstract}


\maketitle

\section{Introduction}

Observation of black holes in the gravitational \cite{LIGOScientific:2016aoc, LIGOScientific:2017vwq, LIGOScientific:2020zkf, Babak:2017tow} and electromagnetic \cite{EventHorizonTelescope:2019dse, Goddi:2016qax} spectra is a promising and intensively developing research area. 
Although it is clear that the least damped fundamental mode is the most important observationally, the first few overtones are important for a number of reasons. 

First, when trying to fit the ringdown signal obtained as a result of the non-linear merger process with a single fundamental quasinormal modes, it was observed that only the late stage of the ringdown could be reproduced with the fundamental mode. However, when the first several overtones are added to the fitting, it is possible to reproduce the whole stage of ring-down \cite{Giesler:2019uxc,Isi:2019aib}. This phenomenon was further studied in \cite{Oshita:2021iyn,Forteza:2021wfq,Oshita:2022pkc} and the possibility of measurement of overtones in future experiments discussed.  

However, the most striking interest to the first few overtones comes from the observation that, unlike the fundamental mode that depends on the geometry near the peak of the effective potential, the overtones bring information on the near-horizon geometry and even slight near-horizon deformation may induce strong deviation of the first few overtones from their Schwarzschild/Kerr values \cite{Konoplya:2023hqb,Konoplya:2022hll, Konoplya:2022pbc}.

Calculation of the higher overtones could be done reliably with the Frobenius method \cite{Leaver:1985ax}. However, for a number of situations it cannot be applied straightforwardly because the wave equation must have polynomial form. After all, for the Frobenius method one needs the initial guess and the mode could easily be missed, if initial guess is not sufficiently close to the true quasinormal frequency. Therefore, the presence of a reliable method that could detect not only the fundamental mode, but also a few first overtones would be highly desirable. The time-domain integration method has not been associated as such method, because for asymptotically flat spacetimes it allows extracting only the fundamental mode with good accuracy \cite{Abdalla:2006fv}. The reason for this, according to our opinion, is related to the fact that at sufficiently late times, the ringing stage is changed by asymptotic power-law tails where quasinormal frequencies are suppressed.

The spectra of black holes in asymptotically de Sitter spacetimes are qualitatively different in this respect: the quasinormal modes dominate in the signal not only at the intermediate, but also at asymptotic times \cite{Dyatlov:2011zz,Dyatlov:2011jd, Konoplya:2024ptj, Konoplya:2022xid}. Therefore, as we will show here, it is possible to extract with unprecedented, for the time-domain integration, accuracy not only the fundamental mode, but also a few higher overtones.
  
Although the spectra of compact objects could hardly depend upon cosmological factors, such as the $\Lambda$-term, which are characterized by large scales and small energy content per unit volume, the quasinormal modes of black hole in various cosmological environment have been actively studied, for example in the Goedel Universe \cite{Konoplya:2005sy}, in the presence of the aether field \cite{Konoplya:2006ar,Ding:2017gfw}, the environment of the galactic halo \cite{Zhao:2023tyo,Figueiredo:2023gas,Konoplya:2022hbl,Zhao:2023itk}  or, most extensively, in the de Sitter world \cite{Molina:2003dc,Jansen:2017oag,Konoplya:2014lha,Zhidenko:2003wq,Konoplya:2007zx,Cuyubamba:2016cug,Dias:2018ynt,Fontana:2018fof,Fontana:2020syy,Konoplya:2013sba}. The latter case has additional interest because there are theories of gravity in which an effective cosmological constant appears without its formal introduction as a cosmological factor.

The paper is structured as follows. In sec. II we describe the black hole and give the wave equations under consideration. Sec. III is devoted to the methods used for the analysis of the spectra, while sec. IV discusses the obtained results. Finally, in the Conclusions we summarize the obtained results. 

\section{The black hole metric and the wave equation}
The metric of the spherically symmetric black hole is given by the following line element,
\begin{equation}\label{metric}
ds^2=-f(r)dt^2+\frac{dr^2}{f(r)}+r^2 (d\theta^2+\sin^2\theta d\phi^2),
\end{equation}
where  the metric function of the Schwazrschild-de Sitter black hole is
\begin{equation}
f(r)= 1 -\frac{2 M}{r} -\frac{\Lambda  r^2}{3},
\end{equation}
Here $M$ is the black hole mass and $\Lambda$ is the cosmological constant.  
The electromagnetic and gravitational perturbations can be reduced to the following  master wave-like equation
\begin{equation}\label{wave-equation}
\dfrac{\partial^2 \Psi}{\partial r_*^2} - \frac{\partial^2 \Psi}{\partial t^2} -V(r) \Psi=0,
\end{equation}
where the ``tortoise coordinate'' $r_*$ has the form:
\begin{equation}
dr_*\equiv\frac{dr}{f(r)}.
\end{equation}
The effective potentials for electromagnetic and axial gravitational perturbations have the following form:
\begin{equation}
V_{e}(r) = f(r) \left(\frac{\ell(\ell+1)}{r^2}\right),
\end{equation}
\begin{equation}
V_{g}(r) = f(r) \left(\frac{\ell(\ell+1)}{r^2}  -\frac{6 M}{r^3} \right)
\end{equation}
where the multipole number ($\ell=1,2,3,...$ for an electromagnetic field and $\ell=2,3,4,...$ for a gravitational one) arises from the separation of the angular variables $\theta$ and $\phi$. The other channel of gravitational perturbations, the polar one, is iso-spectral to the axial. Therefore, here we will consider only one of the gravitational potentials.

\begin{figure*}
\resizebox{\linewidth}{!}{\includegraphics{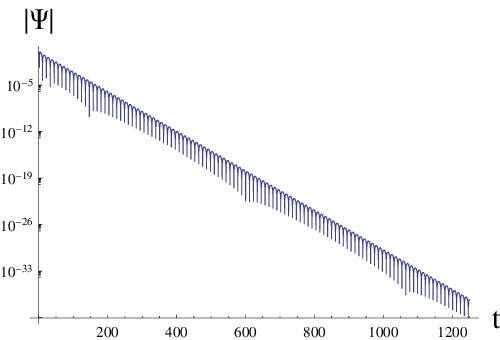}\includegraphics{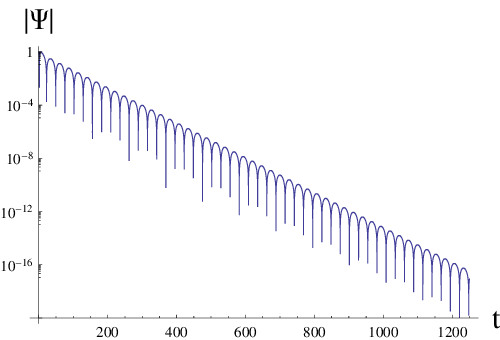}\includegraphics{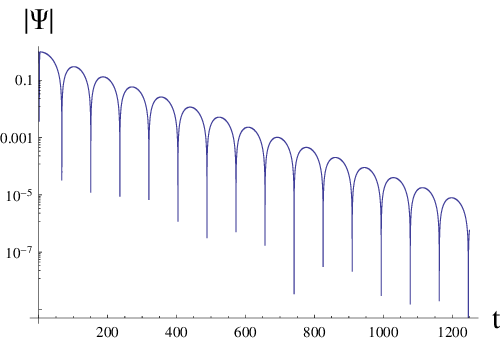}}
\caption{Time-doman profiles for gravitational perturbations $\ell=2$, $\Lambda M^2 =0.1$ (left) and $\Lambda M^2 =0.11$ (right). The integration step is $0.01$ and the fitting is done starting from $t/M = 100$ (left plot), $t/M = 200$ (middle), and $t/M = 330$ (left).}\label{fig:scalarpot}
\end{figure*}

\begin{table*}
\begin{tabular}{|c|c|c|*{4}{c|}c|} 
  \hline
  mode & Frobenius & Time-domain & \multicolumn{4}{c|}{Relative error} \\ \cline{4-7}
    & & & Re & Im &  Mod & Arg \\
   \hline
  $n=0$ & 0.2770303225 - 0.0684424925 i & 0.2770303484  - 0.06844247084 i & 9.35E-08 & 3.16E-07 & 6.99E-08 & 3.94E-07 \\
  $n=1$ & 0.2668830969 - 0.2064740822 i & 0.2668831555 -  0.20647402050 i & 2.2E-07 & 2.99E-07 & 2.55E-08 & 3.81E-07 \\
  $n=2$ & 0.2461936845 - 0.3490193020 i & 0.2461938147 -  0.34901920780 i & 5.29E-07 & 2.7E-07 & 4.51E-09 & 3.93E-07 \\
  $n=3$ & -0.3923379428 i & -0.39233570 i & n/a & 5.72E-06 & 5.72E-06 & n/a \\
  $n=4$ & 0.2173615001 - 0.5015458543 i & 0.2173717281 -  0.50148840771 i & 4.71E-05 & 0.000115 & 8.9E-05 & 5.08E-05 \\
  \hline
\end{tabular}
\caption{First five quasinormal modes for $\Lambda M^2 = 0.05$, $\ell=2$ gravitational perturbations.}\label{table1}
\end{table*}
\begin{table*}
\begin{tabular}{|c|c|c|*{4}{c|}c|}
  \hline
  mode & Frobenius & Time-domain & \multicolumn{4}{c|}{Relative error} \\ \cline{4-7}
    & & & Re & Im &  Mod & Arg \\
   \hline
  $n=0$ & 0.1179164340 - 0.03021048863 i & 0.1179164358 - 0.03021048685 i & 1.53E-08 & 5.89E-08 & 1.07E-08 & 7.11E-08 \\
  $n=1$ & 0.1172432193 - 0.09064094848 i & 0.1172432230 - 0.09064094316 i & 3.16E-08 & 5.87E-08 & 2.2E-09 & 6.64E-08 \\
  $n=2$ & 0.1158764351 - 0.15110181085 i & 0.1158764427 - 0.15110180217 i & 6.56E-08 & 5.76E-08 & 1.2E-08 & 6.49E-08 \\
  $n=3$ & 0.1137720227 - 0.21162078884 i & 0.1137714212 - 0.21162031903 i & 5.29E-06 & 2.22E-06 & 2.91E-06 & 1.19E-06 \\
  $n=4$ & 0.1108542935 - 0.27224084335 i & 0.1109463060 - 0.27247655449 i & 0.00083 & 0.00087 & 0.00086 & 1.05E-05 \\
  \hline
\end{tabular}
\caption{First five quasinormal modes for $\Lambda M^2 = 0.1$, $\ell=2$ gravitational perturbations.}\label{table2}
\end{table*}
\begin{table*}
\begin{tabular}{|c|c|c|*{4}{c|}c|}
  \hline
  mode & Frobenius & Time-domain & \multicolumn{4}{c|}{Relative error} \\ \cline{4-7}
    & & & Re & Im &  Mod & Arg \\
   \hline
  $n=0$ & 0.037269946397 - 0.009615651615 i & 0.037269946132 - 0.009615651475 i & 7.24E-09 & 1.46E-08 & 7.7E-09 & 7.01E-09 \\
  $n=1$ & 0.037249337943 - 0.028846981805 i & 0.037249337739 - 0.028846981388 i & 5.37E-09 & 1.46E-08 & 8.81E-09 & 6.75E-09\\
  $n=2$ & 0.037208059222 - 0.048078393398 i & 0.037208059228 - 0.048078392594 i & 2.69E-10 & 1.68E-08 & 1.04E-08 & 9.08E-09\\
  $n=3$ & 0.037145985562 - 0.067309942430 i & 0.037143274494 - 0.067300357896 i & 7.3E-05 & 0.00014 & 0.00013 & 2.75E-05\\
  $n=4$ & 0.037062927298 - 0.086541687663 i & 0.037047586923 - 0.086575686135 i & 0.00041 & 0.00039 & 0.00027 & 0.00025\\
  \hline
\end{tabular}
\caption{First five quasinormal modes for $\Lambda M^2 = 0.11$, $\ell=2$ gravitational perturbations.}\label{table3}
\end{table*}
\begin{table*}
\begin{tabular}{|c|c|c|*{4}{c|}c|}
  \hline
  mode & Frobenius & Time-domain & \multicolumn{4}{c|}{Relative error} \\ \cline{4-7}
    & & & Re & Im &  Mod & Arg \\
   \hline
  $n=0$ & 0.08034587269 - 0.03027177347 i & 0.08034587341 - 0.03027177243 i & 8.96E-09 & 3.44E-08 & 3.58E-09 & 3.97E-08 \\
  $n=1$ & 0.07926317883 - 0.09084610866 i & 0.07926318119 - 0.09084610563 i & 2.98E-08 & 3.34E-08 & 6.07E-09 & 3.66E-08\\
  $n=2$ & 0.07700867541 - 0.15152843108 i & 0.07700868109 - 0.15152842631 i & 7.38E-08 & 3.17E-08 & 1E-08 & 3.87E-08\\
  $n=3$ & 0.07336386400 - 0.21246336790 i & 0.07336359710 - 0.21246305146 i & 3.64E-06 & 1.49E-06 & 1.72E-06 & 5.35E-07\\
  $n=4$ & 0.06786081944 - 0.27403783819 i & 0.06804888714 - 0.27409403228 i & 0.0028 & 0.00021 & 0.00035 & 0.00045\\
  \hline
\end{tabular}
\caption{First five quasinormal modes for $\Lambda M^2 = 0.1$, $\ell=1$ electromagnetic perturbations.}\label{table4}
\end{table*}

\section{Methods for calculation of quasinormal modes}

Quasinormal modes $\omega_{n}$, where $n$ is the overtone number, is proper oscillation frequency of the black hole, which satisfy the purely ingoing wave at the event horizon and the purely outgoing wave at the de Sitter horizon. Here we briefly summarize the two well-known methods for finding quasinormal modes: time-domain integration and Frobenius method.
For more detailed description of the methods we refer a reader to a review \cite{Konoplya:2011qq}.

\subsection{Time-domain integration}

The tool for an efficient time-domain integration of the master wave equation \ref{wave-equation} was proposed in \cite{Gundlach:1993tp}. 
Equation (\ref{wave-equation}) can be re-written in terms of the light-cone coordinates $du = dt - dr_{*}$ and $dv = dt + dr_{*}$ as follows
\begin{equation}\label{light-cone}
\left(4\frac{\partial^2}{\partial u\partial v}+V(u,v)\right)\Phi(u,v)=0.
\end{equation}
Then, the time evolution operator has the form
\begin{eqnarray}\nonumber
&&\exp\left(h\frac{\partial}{\partial t}\right)=\exp\left(h\frac{\partial}{\partial u}+h\frac{\partial}{\partial v}\right)= \\ \nonumber
&&=\exp\left(h\frac{\partial}{\partial u}\right)+\exp\left(h\frac{\partial}{\partial v}\right) - 1 +
\\\nonumber &&+ \frac{h^2}{2}\left(\exp\left(h\frac{\partial}{\partial u}\right)+\exp\left(h\frac{\partial}{\partial v}\right)\right)\frac{\partial^2}{\partial u\partial v} + \mathcal{O}(h^4).
\end{eqnarray}
Acting by the above operator on the wave function $\Phi$ and using the wave equation (\ref{light-cone}), we obtain
\begin{eqnarray}
\Phi(N)= \Phi(W)+\Phi(E)-\Phi(S) -\nonumber\\
\frac{h^2}{8}V(S)\left(\Phi(W)+\Phi(E)\right) + \mathcal{O}(h^4),\label{integration-scheme}
\end{eqnarray}
where we have: $S=(u,v)$, $W=(u+h,v)$, $E=(u,v+h)$, and $N=(u+h,v+h)$.

The integration scheme given by eq. (\ref{integration-scheme}) enables one to find values of $\Phi$ inside the rhombus constructed on the null-surfaces $u=u_0$ and $v=v_0$, once  the initial data are given on it. As a result we can find the time profile data for the wave function in each point of the rhombus and use the obtained profile for finding quasinormal frequencies.

\subsection{The Prony method}

For extracting the values of frequencies from the time-domain profile we use the Prony method which consists of fitting of the profile data by a sum of damped exponents
\begin{equation}\label{damping-exponents}
\Phi(t)\simeq\sum_{i=1}^pC_ie^{-\imo\omega_i t}.
\end{equation}
We suppose that the quasinormal ringing stage starts at some $t_0=0$ and ends at $t=Nh$, where $N\geq2p-1$. Then, the relation (\ref{damping-exponents}) is satisfied for each point of the profile:
\begin{equation}
x_n\equiv\Phi(nh)=\sum_{j=1}^pC_je^{-\imo\omega_j nh}=\sum_{j=1}^pC_jz_j^n.
\end{equation}

Then we find $z_i$ in terms of the known $x_n$ and calculate the quasinormal frequencies $\omega_i$. For this purpose, a polynomial function $A(z)$ can be fixed as follows
\begin{equation}
A(z)=\prod_{j=1}^p(z-z_j)=\sum_{m=0}^{p}\alpha_m z^{p-m}, \qquad \alpha_0=1.
\end{equation}
Then, using and requirement $\alpha_0=1$ in the following sum
$$\sum_{m=0}^p\alpha_mx_{n-m}=\sum_{m=0}^p\alpha_m\sum_{j=1}^pC_jz_j^{n-m}=$$
$$\sum_{j=1}^pC_jz_j^{n-p}\sum_{m=0}^p\alpha_mz_j^{p-m}=\sum_{j=1}^pC_jz_j^{n-p}A(z_j)=0,$$
we obtain 
\begin{equation}\label{Prony-equation}
\sum_{m=1}^p\alpha_mx_{n-m}=-x_{n}.
\end{equation}
Using $n=p,...,N$ in Eq.~(\ref{Prony-equation}) one can find $N-p+1\geq p$ linear equations for $\alpha_m$.

The above equation can be cast in the following matrix form
$$\left(\begin{array}{llll}
    x_{p-1} & x_{p-2} & \ldots & x_0 \\
    x_p & x_{p-1} & \ldots & x_1 \\
    \vdots & \vdots & \ddots & \vdots \\
    x_{N-1} & x_{N-2} & \ldots & x_{N-p} \\
  \end{array}\right)
\left(\begin{array}{c}
    \alpha_1 \\
    \alpha_2 \\
    \vdots \\
    \alpha_p\\
\end{array}\right)=-
\left(\begin{array}{c}
    x_p \\
    x_{p+1} \\
    \vdots \\
    x_N\\
\end{array}\right).
$$
The above matrix equation can be solved as follows
\begin{equation}
\alpha=-(X^+X)^{-1}X^+x,
\end{equation}
where $X^+$ is a Hermitian transposition of $X$. Once $\alpha_m$ are found, one can find numerically the roots $z_j$ and the quasinormal frequencies $\omega_j$:
$$\omega_j=\frac{\imo}{h}\ln(z_j).$$

The time-domain integration method together with the Prony method allows one to find the fundamental mode and test stability of black holes. It was effectively used in a great number of works (see, for instance, \cite{Abdalla:2006qj, Ishihara:2008re, Bronnikov:2021liv, Bolokhov:2023ruj, Konoplya:2020jgt, Skvortsova:2023zca, Churilova:2019qph} and references therein), though only to determine the fundamental mode. Nevertheless, some recent calculations done in \cite{Skvortsova:2023zmj, Dubinsky:2024hmn, Konoplya:2024ptj} for the asymptotically de Sitter metric allowed to use it also for an accurate finding of the first the overtone as well.

\subsection{Frobenius method}

The Frobenius method \cite{ Leaver:1985ax} allows us to calculate the quasinormal frequencies precisely, provided we have a sufficiently close initial guess for the frequency. 
The method is based on the convergent routine and is related in detail in a great number of works. For the particular case of the Schwarzschild-de Sitter black holes it was related in  \cite{Konoplya:2004uk}. Here we mention the milestones of the method. 

The solution of the wave equation is expanded into the Frobenius series 
\begin{widetext}
\be\label{Frobenius} 
\P(r^*)=\left(\frac 1 r - \frac 1 {r_e}\right)^{\rho_e}\left(\frac 1
{r_c}-\frac 1 r\right)^{-\rho_c}\left(\frac 1 r+\frac 1
{r_e+r_c}\right)^{\rho_c+\rho_e}\sum_{n\geq0}a_n\left(\frac{\frac
1 r-\frac1{r_e}}{\frac1{r_c}-\frac1{r_e}}\right)^n, \ee 
where
$r_e$ is the event horizon, $r_c$ is the cosmological horizon,
$\rho_e$ and $\rho_c$ have the following form
$$\rho_e=\frac{\imo\om}{\displaystyle 2M\left(\frac{1}{r_c}-\frac{1}{r_e}\right)\left(\frac{1}{r_c+r_e}+\frac{1}{r_e}\right)};\qquad\rho_c=\frac{-\imo\om}{\displaystyle 2M\left(\frac{1}{r_c}-\frac{1}{r_e}\right)\left(\frac{1}{r_c+r_e}+\frac{1}{r_c}\right)}.$$
\end{widetext}
Using the above expansion into the wave-like equation, we reduce the problem to the three terms recurrence relation:
for $a_n$ \be\label{reccur}
a_{n+1}\a_n+a_n\b_n+a_{n-1}\g_n=0, \qquad n\geq0, \quad \g_0=0.
\ee
Once the coefficients of the expansion are found, the frequencies can be obtained as roots of the corresponding algebraic equation.  
For quicker convergence we also used the Nollert improvement \cite{Nollert:1993zz,Zhidenko:2006rs} of the Leaver procedure \cite{Leaver:1985ax}. 
The Frobenius method has been actively used in a numerous publications as a reliable tool for finding precise values of quasinormal frequencies including higher overtones \cite{Zinhailo:2024jzt,Nollert:1993zz,Hirano:2024fgp,Kodama:2009bf,Bolokhov:2023bwm}.

\section{Quasinormal modes via time-domain and Frobenius methods}

The time domain integration method was extensively studied as to the opportunity to extract higher modes from the time domain profiles of asymptotically flat (Schwarzschild) black holes \cite{Abdalla:2006fv} and the conclusion was that while the fundamental mode could be studied for $\ell =1$ and $\ell=2$ perturbations with reasonable accuracy, already the first overtones could only be roughly estimated. The larger $\ell$ is, the longer is the ringing period and the higher $n$ could be detected, so that the second overtone could be estimated only for $\ell \geq 8$. This means that in the asymptotically flat spacetimes time-domain integration method is not of much value, because at such large $\ell$ the automatic WKB method gives quite a few overtones with remarkable accuracy \cite{Konoplya:2019hlu}. Thus, overall, for the lowest multipole numbers, which is of our primary interest in perturbation theory, the time-domain integration method could only find the fundamental mode and sometimes give a rough estimate for the first overtone.

Precise values of quasinormal modes including first few overtones for the gravitational and electromagentic perturbations calculated with the Frobenius method could be found in \cite{Konoplya:2004uk}. Nevertheless, we calculated them here independently, because of the choice of slightly different parameters of the system.

The least damped mode $n=9$, as can be seen in Table I-IV, could be extracted from the time-domain profiles with an unprecedented accuracy for this method of seven-eight digits. Even the fourth overtone is obtained with a relative error consisting, at the maximum, of a small fraction of one percent. Here we integrated the wave equation with a step $\delta =0.01$ and diminishing the integration step and extending the time of integration allows finding higher frequencies.

\section{Conclusions}

This work was devoted to a technical aspect of the calculation of quasinormal modes, which, nevertheless, may have interesting applications.
When it is necessary to calculate the first few overtones for the lowest multipole numbers, such as $\ell=2$ of gravitational perturbations or $\ell=1$ electromagentic ones, only some methods are applicable for this problem. The popular WKB approach does not work well for $\ell <n$ and, moreover, converges only asymptotically, so that its accuracy  strictly speaking is unknown and not guaranteed. The Frobenius method is accurate, but requires a strictly polynomial form of the wave equation and may be cumbersome, leading to many terms recurrence relations. After all, it requires an initial guess of the frequency, so that sometimes modes could be simply missed. The convergence of the other methods, such as Bernstein polynomial method or shooting, also may be not good for complex wave equations. In this situation, it could be useful to have an automatic method (i.e. the one which does not require modifications for different metrics and effective potentials) which could reliably detect the few several overtones. Quite unexpectedly, we propose the time domain integration for this task. For an asymptotically flat black holes the time-domain integration usually does not allow to detect higher modes \cite{Abdalla:2006fv}. Nevertheless, if the black hole is asymptotically flat, the strategy then would be to attribute tiny value of the cosmological constant to the metric, simply by adding the de Sitter term $\sim r^2$ to $f(r)$, and then calculate the frequencies with the time-domain integration. As the power-law tails are absent for such system and quasinormal modes governs the signal at asymptotically late times as well, this strategy could work for asymptotically flat black holes as well and could be performed with a personal computer during reasonable time.

\acknowledgments
The author thanks R.A. Konoplya for useful discussions. The author acknowledges the University of Seville for its support through the Plan-US of aid to Ukraine.

\bibliographystyle{unsrt}
\bibliography{Bibliography}

\begin{thebibliography}{10}

\bibitem{LIGOScientific:2016aoc}
B.~P. Abbott et~al.
\newblock {Observation of Gravitational Waves from a Binary Black Hole Merger}.
\newblock {\em Phys. Rev. Lett.}, 116(6):061102, 2016.

\bibitem{LIGOScientific:2017vwq}
B.~P. Abbott et~al.
\newblock {GW170817: Observation of Gravitational Waves from a Binary Neutron Star Inspiral}.
\newblock {\em Phys. Rev. Lett.}, 119(16):161101, 2017.

\bibitem{LIGOScientific:2020zkf}
R.~Abbott et~al.
\newblock {GW190814: Gravitational Waves from the Coalescence of a 23 Solar Mass Black Hole with a 2.6 Solar Mass Compact Object}.
\newblock {\em Astrophys. J. Lett.}, 896(2):L44, 2020.

\bibitem{Babak:2017tow}
Stanislav Babak, Jonathan Gair, Alberto Sesana, Enrico Barausse, Carlos~F. Sopuerta, Christopher P.~L. Berry, Emanuele Berti, Pau Amaro-Seoane, Antoine Petiteau, and Antoine Klein.
\newblock {Science with the space-based interferometer LISA. V: Extreme mass-ratio inspirals}.
\newblock {\em Phys. Rev. D}, 95(10):103012, 2017.

\bibitem{EventHorizonTelescope:2019dse}
Kazunori Akiyama et~al.
\newblock {First M87 Event Horizon Telescope Results. I. The Shadow of the Supermassive Black Hole}.
\newblock {\em Astrophys. J. Lett.}, 875:L1, 2019.

\bibitem{Goddi:2016qax}
C.~Goddi et~al.
\newblock {BlackHoleCam: Fundamental physics of the galactic center}.
\newblock {\em Int. J. Mod. Phys. D}, 26(02):1730001, 2016.

\bibitem{Giesler:2019uxc}
Matthew Giesler, Maximiliano Isi, Mark~A. Scheel, and Saul Teukolsky.
\newblock {Black Hole Ringdown: The Importance of Overtones}.
\newblock {\em Phys. Rev. X}, 9(4):041060, 2019.

\bibitem{Isi:2019aib}
Maximiliano Isi, Matthew Giesler, Will~M. Farr, Mark~A. Scheel, and Saul~A. Teukolsky.
\newblock {Testing the no-hair theorem with GW150914}.
\newblock {\em Phys. Rev. Lett.}, 123(11):111102, 2019.

\bibitem{Oshita:2021iyn}
Naritaka Oshita.
\newblock {Ease of excitation of black hole ringing: Quantifying the importance of overtones by the excitation factors}.
\newblock {\em Phys. Rev. D}, 104(12):124032, 2021.

\bibitem{Forteza:2021wfq}
Xisco~Jim\'enez Forteza and Pierre Mourier.
\newblock {High-overtone fits to numerical relativity ringdowns: Beyond the dismissed n=8 special tone}.
\newblock {\em Phys. Rev. D}, 104(12):124072, 2021.

\bibitem{Oshita:2022pkc}
Naritaka Oshita.
\newblock {Thermal ringdown of a Kerr black hole: overtone excitation, Fermi-Dirac statistics and greybody factor}.
\newblock {\em JCAP}, 04:013, 2023.

\bibitem{Konoplya:2023hqb}
R.~A. Konoplya.
\newblock {The sound of the event horizon}.
\newblock {\em Int. J. Mod. Phys. D}, 32(14):2342014, 2023.

\bibitem{Konoplya:2022hll}
R.~A. Konoplya, A.~F. Zinhailo, J.~Kunz, Z.~Stuchlik, and A.~Zhidenko.
\newblock {Quasinormal ringing of regular black holes in asymptotically safe gravity: the importance of overtones}.
\newblock {\em JCAP}, 10:091, 2022.

\bibitem{Konoplya:2022pbc}
R.~A. Konoplya and A.~Zhidenko.
\newblock {First few overtones probe the event horizon geometry, arXiv: 2209.00679}.
\newblock 9 2022.

\bibitem{Leaver:1985ax}
E.~W. Leaver.
\newblock {An Analytic representation for the quasi normal modes of Kerr black holes}.
\newblock {\em Proc. Roy. Soc. Lond. A}, 402:285--298, 1985.

\bibitem{Abdalla:2006fv}
E.~Abdalla and D.~Giugno.
\newblock {An Extensive search for overtones in Schwarzschild black holes}.
\newblock {\em Braz. J. Phys.}, 37:450--456, 2007.

\bibitem{Dyatlov:2011zz}
Semyon Dyatlov.
\newblock {Exponential energy decay for Kerr-de Sitter black holes beyond event horizons}.
\newblock {\em Math. Res. Lett.}, 18:1023--1035, 2011.

\bibitem{Dyatlov:2011jd}
Semyon Dyatlov.
\newblock {Asymptotic distribution of quasi-normal modes for Kerr-de Sitter black holes}.
\newblock {\em Annales Henri Poincare}, 13:1101--1166, 2012.

\bibitem{Konoplya:2024ptj}
R.~A. Konoplya.
\newblock {Two Regimes of Asymptotic Fall-off of a Massive Scalar Field in the Schwarzschild-de Sitter Spacetime, arXiv: 2401.17106}.
\newblock 1 2024.

\bibitem{Konoplya:2022xid}
R.~A. Konoplya and A.~Zhidenko.
\newblock {Nonoscillatory gravitational quasinormal modes and telling tails for Schwarzschild\textendash{}de Sitter black holes}.
\newblock {\em Phys. Rev. D}, 106(12):124004, 2022.

\bibitem{Konoplya:2005sy}
R.~A. Konoplya and Elcio Abdalla.
\newblock {Scalar field perturbations of the Schwarzschild black hole in the Godel universe}.
\newblock {\em Phys. Rev. D}, 71:084015, 2005.

\bibitem{Konoplya:2006ar}
R.~A. Konoplya and A.~Zhidenko.
\newblock {Gravitational spectrum of black holes in the Einstein-Aether theory}.
\newblock {\em Phys. Lett. B}, 648:236--239, 2007.

\bibitem{Ding:2017gfw}
Chikun Ding.
\newblock {Quasinormal ringing of black holes in Einstein-aether theory}.
\newblock {\em Phys. Rev. D}, 96(10):104021, 2017.

\bibitem{Zhao:2023tyo}
Yuqian Zhao, Bing Sun, Kai Lin, and Zhoujian Cao.
\newblock {Axial gravitational ringing of a spherically symmetric black hole surrounded by dark matter spike}.
\newblock {\em Phys. Rev. D}, 108(2):024070, 2023.

\bibitem{Figueiredo:2023gas}
Enzo Figueiredo, Andrea Maselli, and Vitor Cardoso.
\newblock {Black holes surrounded by generic dark matter profiles: Appearance and gravitational-wave emission}.
\newblock {\em Phys. Rev. D}, 107(10):104033, 2023.

\bibitem{Konoplya:2022hbl}
R.~A. Konoplya and A.~Zhidenko.
\newblock {Solutions of the Einstein Equations for a Black Hole Surrounded by a Galactic Halo}.
\newblock {\em Astrophys. J.}, 933(2):166, 2022.

\bibitem{Zhao:2023itk}
Yuqian Zhao, Bing Sun, Zhoujian Cao, Kai Lin, and Wei-Liang Qian.
\newblock {Influence of dark matter equation of state on the axial gravitational ringing of supermassive black holes}.
\newblock {\em Phys. Rev. D}, 109(4):044031, 2024.

\bibitem{Molina:2003dc}
C.~Molina, D.~Giugno, E.~Abdalla, and A.~Saa.
\newblock {Field propagation in de Sitter black holes}.
\newblock {\em Phys. Rev. D}, 69:104013, 2004.

\bibitem{Jansen:2017oag}
Aron Jansen.
\newblock {Overdamped modes in Schwarzschild-de Sitter and a Mathematica package for the numerical computation of quasinormal modes}.
\newblock {\em Eur. Phys. J. Plus}, 132(12):546, 2017.

\bibitem{Konoplya:2014lha}
R.~A. Konoplya and A.~Zhidenko.
\newblock {Charged scalar field instability between the event and cosmological horizons}.
\newblock {\em Phys. Rev. D}, 90(6):064048, 2014.

\bibitem{Zhidenko:2003wq}
A.~Zhidenko.
\newblock {Quasinormal modes of Schwarzschild de Sitter black holes}.
\newblock {\em Class. Quant. Grav.}, 21:273--280, 2004.

\bibitem{Konoplya:2007zx}
R.~A. Konoplya and A.~Zhidenko.
\newblock {Decay of a charged scalar and Dirac fields in the Kerr-Newman-de Sitter background}.
\newblock {\em Phys. Rev. D}, 76(8):084018, 2007.
\newblock [Erratum: Phys.Rev.D 90, 029901 (2014)].

\bibitem{Cuyubamba:2016cug}
M.~A. Cuyubamba, R.~A. Konoplya, and A.~Zhidenko.
\newblock {Quasinormal modes and a new instability of Einstein-Gauss-Bonnet black holes in the de Sitter world}.
\newblock {\em Phys. Rev. D}, 93(10):104053, 2016.

\bibitem{Dias:2018ynt}
Oscar J.~C. Dias, Felicity~C. Eperon, Harvey~S. Reall, and Jorge~E. Santos.
\newblock {Strong cosmic censorship in de Sitter space}.
\newblock {\em Phys. Rev. D}, 97(10):104060, 2018.

\bibitem{Fontana:2018fof}
R.~D.~B. Fontana, Jeferson de~Oliveira, and A.~B. Pavan.
\newblock {Dynamical evolution of non-minimally coupled scalar field in spherically symmetric de Sitter spacetimes}.
\newblock {\em Eur. Phys. J. C}, 79(4):338, 2019.

\bibitem{Fontana:2020syy}
R.~D.~B. Fontana, P.~A. Gonz\'alez, Eleftherios Papantonopoulos, and Yerko V\'asquez.
\newblock {Anomalous decay rate of quasinormal modes in Reissner-Nordstr\"om black holes}.
\newblock {\em Phys. Rev. D}, 103(6):064005, 2021.

\bibitem{Konoplya:2013sba}
R.~A. Konoplya and A.~Zhidenko.
\newblock {Instability of D-dimensional extremally charged Reissner-Nordstrom(-de Sitter) black holes: Extrapolation to arbitrary D}.
\newblock {\em Phys. Rev. D}, 89(2):024011, 2014.

\bibitem{Konoplya:2011qq}
R.~A. Konoplya and A.~Zhidenko.
\newblock {Quasinormal modes of black holes: From astrophysics to string theory}.
\newblock {\em Rev. Mod. Phys.}, 83:793--836, 2011.

\bibitem{Gundlach:1993tp}
Carsten Gundlach, Richard~H. Price, and Jorge Pullin.
\newblock {Late time behavior of stellar collapse and explosions: 1. Linearized perturbations}.
\newblock {\em Phys. Rev. D}, 49:883--889, 1994.

\bibitem{Abdalla:2006qj}
E.~Abdalla, B.~Cuadros-Melgar, A.~B. Pavan, and C.~Molina.
\newblock {Stability and thermodynamics of brane black holes}.
\newblock {\em Nucl. Phys. B}, 752:40--59, 2006.

\bibitem{Ishihara:2008re}
Hideki Ishihara, Masashi Kimura, Roman~A. Konoplya, Keiju Murata, Jiro Soda, and Alexander Zhidenko.
\newblock {Evolution of perturbations of squashed Kaluza-Klein black holes: escape from instability}.
\newblock {\em Phys. Rev. D}, 77:084019, 2008.

\bibitem{Bronnikov:2021liv}
Kirill~A. Bronnikov, Roman~A. Konoplya, and Thomas~D. Pappas.
\newblock {General parametrization of wormhole spacetimes and its application to shadows and quasinormal modes}.
\newblock {\em Phys. Rev. D}, 103(12):124062, 2021.

\bibitem{Bolokhov:2023ruj}
S.~V. Bolokhov.
\newblock {Long-lived quasinormal modes and oscillatory tails of the Bardeen spacetime}.
\newblock {\em Phys. Rev. D}, 109(6):064017, 2024.

\bibitem{Konoplya:2020jgt}
R.~A. Konoplya, A.~F. Zinhailo, and Z.~Stuchlik.
\newblock {Quasinormal modes and Hawking radiation of black holes in cubic gravity}.
\newblock {\em Phys. Rev. D}, 102(4):044023, 2020.

\bibitem{Skvortsova:2023zca}
Milena Skvortsova.
\newblock {Stability of Asymptotically Flat $\mathbf{(2+1)}$-Dimensional Black Holes with Gauss\textendash{}Bonnet Corrections}.
\newblock {\em Grav. Cosmol.}, 30(1):68--70, 2024.

\bibitem{Churilova:2019qph}
M.~S. Churilova, R.~A. Konoplya, and A.~Zhidenko.
\newblock {Arbitrarily long-lived quasinormal modes in a wormhole background}.
\newblock {\em Phys. Lett. B}, 802:135207, 2020.

\bibitem{Skvortsova:2023zmj}
Milena Skvortsova.
\newblock {Quasinormal spectrum of $(2+1)$-dimensional asymptotically flat, dS and AdS black holes}.
\newblock 11 2023.

\bibitem{Dubinsky:2024hmn}
Alexey Dubinsky and Antonina Zinhailo.
\newblock {Asymptotic decay and quasinormal frequencies of scalar and Dirac fields around dilaton-de Sitter black holes, arXiv: 2404.01834}.
\newblock 4 2024.

\bibitem{Konoplya:2004uk}
R.~A. Konoplya and A.~Zhidenko.
\newblock {High overtones of Schwarzschild-de Sitter quasinormal spectrum}.
\newblock {\em JHEP}, 06:037, 2004.

\bibitem{Nollert:1993zz}
Hans-Peter Nollert.
\newblock {Quasinormal modes of Schwarzschild black holes: The determination of quasinormal frequencies with very large imaginary parts}.
\newblock {\em Phys. Rev. D}, 47:5253--5258, 1993.

\bibitem{Zhidenko:2006rs}
Alexander Zhidenko.
\newblock {Massive scalar field quasi-normal modes of higher dimensional black holes}.
\newblock {\em Phys. Rev. D}, 74:064017, 2006.

\bibitem{Zinhailo:2024jzt}
Antonina~F. Zinhailo.
\newblock {Exploring Unique Quasinormal Modes of a Massive Scalar Field in Brane-World Scenarios}.
\newblock 3 2024.

\bibitem{Hirano:2024fgp}
Shin'ichi Hirano, Masashi Kimura, Masahide Yamaguchi, and Jiale Zhang.
\newblock {Parametrized Black Hole Quasinormal Ringdown Formalism for Higher Overtones}.
\newblock 4 2024.

\bibitem{Kodama:2009bf}
Hideo Kodama, R.~A. Konoplya, and Alexander Zhidenko.
\newblock {Gravitational stability of simply rotating Myers-Perry black holes: Tensorial perturbations}.
\newblock {\em Phys. Rev. D}, 81:044007, 2010.

\bibitem{Bolokhov:2023bwm}
S.~V. Bolokhov.
\newblock {Long-lived quasinormal modes and overtones' behavior of the holonomy corrected black holes}.
\newblock 11 2023.

\bibitem{Konoplya:2019hlu}
R.~A. Konoplya, A.~Zhidenko, and A.~F. Zinhailo.
\newblock {Higher order WKB formula for quasinormal modes and grey-body factors: recipes for quick and accurate calculations}.
\newblock {\em Class. Quant. Grav.}, 36:155002, 2019.

\end{thebibliography}

\end{document}